# MEECDA: Multihop Energy Efficient Clustering and Data Aggregation Protocol for HWSN

Surender Kumar
University of Petroleum and Energy Studies, India

M. Prateek, N.J. Ahuja
University of Petroleum and Energy Studies, India

Bharat Bhushan
Guru Nanak Khalsa College
Yamunanagar, India

## ABSTRACT
Wireless sensor network consists of large number of inexpensive tiny sensors which are connected with low power wireless communications. Most of the routing and data dissemination protocols of WSN assume a homogeneous network architecture, in which all sensors have the same capabilities in terms of battery power, communication, sensing, storage, and processing. However the continued advances in miniaturization of processors and low-power communications have enabled the development of a wide variety of nodes. When more than one type of node is integrated into a WSN, it is called heterogeneous. Multihop short distance communication is an important scheme to reduce the energy consumption in a sensor network because nodes are densely deployed in a WSN. In this paper M-EECDA (Multihop Energy Efficient Clustering & Data Aggregation Protocol for Heterogeneous WSN) is proposed and analyzed. The protocol combines the idea of multihop communications and clustering for achieving the best performance in terms of network life and energy consumption. M-EECDA introduces a sleep state and three tier architecture for some cluster heads to save energy of the network. M-EECDA consists of three types of sensor nodes: normal, advance and super. To become cluster head in a round normal nodes use residual energy based scheme. Advance and super nodes further act as relay node to reduce the transmission load of a normal node cluster head when they are not cluster heads in a round.

## 1. INTRODUCTION
Technological advances of electromechanical devices help in the development of tiny sensors with sensing, processing and communication capabilities [1] [3] [4] [6]. WSNs have large number of tiny sensors that are deployed densely inside or very close to an area for monitoring it [1]. Sensor nodes have the ability of self organization [2] and after collecting the data from sensing field they send it to a sink or base station for further processing via radio communication.

Sensor networks are designed mainly for unattended operation and in most of the applications nodes generally remain stationary after deployment. Quality of service is a major concern in traditional networks however energy consumption is an important issue in sensor networks [1] [6]. Since nodes are battery operated devices and changing or recharging of battery is almost impossible due to the large number and harsh environment deployment of these tiny sensors. Due to dense deployment of sensor nodes cluster based routing is an important scheme for reducing energy consumption and prolonging the life of a sensor network [4]. In a cluster based routing, communications between nodes are controlled by cluster heads. Cluster head is either randomly selected from nodes by using a probability scheme or a centralized control algorithm is used for this purpose so that the energy load can be evenly distributed in the network [5]. Cluster head is responsible to collect the data from its members, aggregates it and then send it to a sink or base station. Thus a two layer routing scheme is used in cluster based routing, first layer is used to select the cluster head from sensor nodes and the second layer is used for routing [4].

There are three common types of resource heterogeneity in sensor nodes: computational heterogeneity, link heterogeneity and energy heterogeneity [19]. Heterogeneity in WSNs helps in improving the scalability and non uniform energy drainage of the network. In this research paper multihop energy efficient clustering and data aggregation protocol for heterogeneous network is proposed. The protocol suggests a new scheme for the cluster head selection on the basis of initial and residual energy of nodes. It will further use the multihop short distance communications for reducing the energy consumption and improve the life of a sensor network. Rest of this paper is organized in to the section as follows. Section 2 describes the related work of cluster based protocols, section 3 explains the wireless sensor network model and section 4 describes the radio energy model which has been used in this paper, section 5 describes the proposed M-EECDA protocol, section 6 shows simulation results and finally the paper is concluded in section 7.

## 2. RELATED WORK
Traditional wireless network protocols of cellular and mobile adhoc network cannot be applied directly in wireless sensor network because they do not consider the energy, processing and storage constraints of sensor network [1][6]. In the recent years many new cluster based routing protocols have been designed for sensor network. LEACH [5] is probably the first and an important protocol in this category. LEACH randomly selects the few sensor nodes to perform the duty of a cluster head so that energy load can be evenly distribute among the sensor nodes. LEACH operation is divided into rounds and each round further consists of two phase: setup and steady state. Cluster head election procedure is performed in setup phase. The aggregation of collected data and further transmission to the base station is completed in steady state phase. To avoid the intracluster and intercluster collision LEACH uses a CSMA/TDMA based scheme. Cluster head is elected by generating a random number between 0 and 1 and if this number is less than a particular threshold value T (n) then the node becomes cluster head for the current round. Author of [5] through simulation shows that only 5 percent of the nodes need to become a cluster head in a round.

$$T(n) = \begin{cases} \frac{p}{1-p \times (r \bmod \frac{1}{p})} & \text{If } n \in G \\ 0 & \text{otherwise} \end{cases} \quad (1)$$

Here G denotes the set of nodes that are not selected as a cluster head in last $\frac{1}{p}$ rounds and r is the current round.

In [5] another protocol LEACH-C is proposed which uses a base station controlled centralized algorithm for the selection of cluster head. LEACH-C has the steady state phase similar to LEACH but different setup phase. Initially in setup phase each node sends the information about their location and energy level to the base station by using a GPS. Base station computes the average energy of the sensor network and the nodes which have energy below this cannot become cluster head for the current round. From the remaining nodes base station finds out the best node for cluster head election.

PEGASIS [7] is an optimal chain based protocol where each node transmits and receives only from its close neighbors through a chain. The chain can be formed either by the nodes itself using a greedy algorithm or the base station computes the chain and then broadcasts it to the nodes. The aggregated data can be sent to base station by any node of the chain and the nodes will take turns for sending the data to base station.

HEED [8] is proposed for homogeneous network which uses a hybrid approach for cluster head election. Probability scheme based on the residual energy is used for cluster head election and nodes join a particular cluster for which they have the minimum communication cost.

TEEN is a reactive protocol for time critical applications where cluster head broadcasts two values known as hard and soft threshold [9]. Hard threshold value is for sensed attributes and soft is for small changes in the sensed value. Nodes continuously sense the region but data transmission is done less frequently. During a round nodes switch on its transmitter only when the sensed value is more than the hard threshold. This value is stored in an intermediate variable and in the current round nodes will transmit the data again if both of the following conditions are satisfied.

- Sensed attribute current value is more than the hard threshold.
- Sensed attribute differs from intermediate variable by a value which is greater than or equal to soft threshold.

Thus hard threshold reduces the no of transmissions by transmitting only when the sensed value is in the region of interest and soft value further reduces the transmission when there is a little or no change in the sensed value.

APTEEN is another protocol proposed for time critical applications in which periodicity and threshold value is changed according to user needs and application type [10]. APTEEN is equally good for reactive and proactive policies. Additional complexity of thresholds and count time is the disadvantage of this scheme. In [11] author proposes a distributed, randomized algorithm for organizing the sensor network into cluster and the algorithm is further extended to generate a hierarchy of cluster heads for saving the energy.

SEP [12] study the impact of heterogeneity on cluster based protocol. A heterogeneity source is created in the sensor network by introducing some nodes of high energy and weighted election probabilities of the nodes are used for the election of cluster head according to the remaining energy of the nodes. Threshold for normal nodes and advanced nodes are calculated on the basis of weighted probabilities of normal and advance nodes.

DEEC is a distributed energy efficient protocol for heterogeneous network in which nodes become cluster head on the basis of residual and average energy of the network. Energy expenditure of nodes is controlled by means of an adaptive approach and for this average energy is used as the reference energy [13]. Author of [14] analyses the strengths and weakness of existing and new protocols.

DBCP [15] is a protocol for three level heterogeneous networks which uses a new cluster head election scheme based on the initial energy and average distance of the nodes from the sink. EEPSC [16] divides the network into many static clusters and cluster head distributes the energy load of the network among high power sensors to prolong the network lifetime.

EEHCA uses a backup cluster head for improving the performance of the network and uses a method for the cluster head election which is based on the distance of the nodes and center of the cluster head [17]. Backup cluster head works when the energy of the primary cluster head is depleted.

EECDA [18] proposes a three level heterogeneous model where some percentages of the nodes have more energy than the normal nodes which are known as advanced nodes. Further in advanced nodes some fractions of the nodes have even more energy than the normal nodes which are known as super nodes. EECDA is a protocol for heterogeneous WSN which improves the network lifetime and stability by combining cluster based routing and data aggregation techniques. Novel cluster head election scheme and a path of maximum sum of energy residues for data transmission are used in EECDA for achieving these objectives.

# 3. NETWORK MODEL

This section contains the description of the wireless sensor network model which is used for this research paper. Model contains n sensor nodes which are randomly deployed in a 100 x 100 square meters region as shown in Figure 1. Various assumptions make about the network model and nodes are as follows.

- Nodes continuously sense the region and they always have the data for sensing to base station.
- Nodes are deployed randomly in the region.
- Base station and nodes become stationary after deployment.
- Base station is located in the middle of sensing region.
- Nodes are location unaware i.e. they do not have any information about their location.
- Battery of the nodes cannot be changed or recharged because they are densely deployed in a harsh environment.

- In the sensing region there are three types of sensor nodes i.e. super, advanced and normal nodes. Super and advanced nodes have more energy than the normal nodes.

In the cluster based protocol cluster head does the communication with its members and also the aggregation of collected data to save the energy. If n sensors are distributed uniformly in M x M region and k is the optimal number of clusters per round. Then on average $\frac{n}{k}$ nodes are per cluster (one cluster head and $(\frac{n}{k}-1)$ non-cluster head nodes). Base station is located in the mid of sensing region and according to [5] the energy dissipated by a cluster head during a single round can be found by using the equation (2)

$$E_{CH} = \left(\frac{n}{k}-1\right) L.E_{elec} + \frac{n}{k} L.E_{DA} + L.E_{elec} + L.\varepsilon_{fs} d_{BS}^2 \quad (2)$$

Where L is the no of bits of the data message, $d_{BS}$ is the distance between base station and cluster head, $E_{DA}$ is the energy required to perform data aggregation in a round and k is the number of clusters. Since cluster members transmit data to only its cluster head therefore energy dissipated in a non cluster head follows the free space path model ($d^2$ power loss) and it is given by Equation (3).

$$E_{NCH} = L.E_{elec} + L.\varepsilon_{fs} d_{CH}^2 \quad (3)$$

Here $d_{CH}$ is the distance between nodes and cluster head. The energy depleted in a cluster is given by Equation (4).

$$E_{cluster} \approx E_{CH} + \frac{n}{k} E_{NCH} \quad (4)$$

The total energy which is dissipated in the network will be:

$$E_{total} = L(2nE_{elec} + nE_{DA} + \varepsilon_{fs}(kd_{BS}^2 + nd_{CH}^2)) \quad (5)$$

The optimal number of clusters can be found by finding the derivative of $E_{total}$ with respect to k and equating it to zero.

$$k_{opt} = \frac{\sqrt{n}}{\sqrt{2\pi}} \sqrt{\frac{\epsilon_{fs}}{\epsilon_{mp}} \frac{M}{d_{BS}^2}} \quad (6)$$

$p_{opt}$ is the optimal probability of a node to become cluster head and it can be calculated as follows

$$p_{opt} = \frac{k_{opt}}{n} \quad (7)$$

## 4. RADIO ENERGY MODEL

Radio dissipation energy model of [5] is used in this paper. There are three parts of radio model: transmitter, power amplifier and the receiver. Two propagation models: free space ($d^2$ power loss) and multipath fading ($d^4$ power loss) channel are used. The energy spent for transmitting an L bit data message to a distance d is given by Equation (8)

$$E_{TX(L,d)} = \begin{pmatrix} L \times E_{elec} + \epsilon_{fs} \times d^2 & if\ d < d_0 \\ L \times E_{elec} + \epsilon_{mp} \times d^4 & if\ d \geq d_0 \end{pmatrix} \quad (8)$$

$E_{elec}$ is the electricity spent to run the transmitter or receiver circuitry. The parameters $\epsilon_{mp}$ and $\epsilon_{fs}$ denotes the amount of energy spent per bit in the radio frequency amplifier according to the cross over distance $d_0$ which is given by Equation (9).

$$d_0 = \sqrt{\frac{\epsilon_{fs}}{\epsilon_{mp}}} \quad (9)$$

The energy expanded to receive an L bit message is given by Equation (10).

$$E_{RX(L,d)} = L \times E_{elec} \quad (10)$$

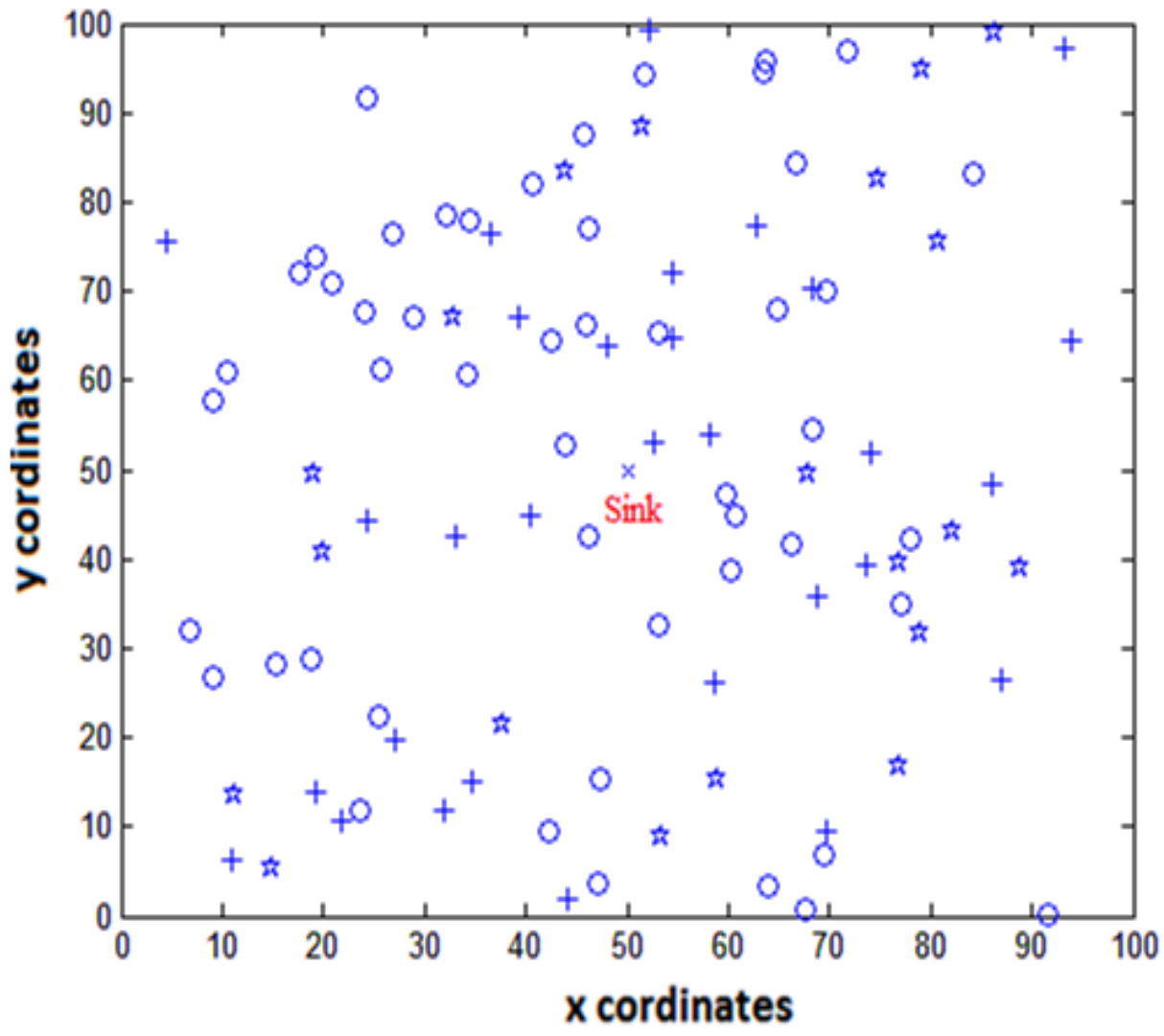

**Figure 1 Sensing Region**

**(o – Normal, + - Advance,* - Super Node, x -Sink)**

## 5. M-EECDA

M-EECDA (Multihop Energy Efficient Clustering and Data Aggregation Protocol for Heterogeneous WSN) main aim is to maintain the energy consumption of the network efficiently. It saves the energy of the system with the introduction of multihop communication for normal cluster head. M-EECDA consists of three types of sensor nodes (i.e. normal, advanced and super) which are randomly deployed in a sensing region. Let m be the fraction of normal nodes among total nodes N nodes, and $m_0$ is the fraction of super nodes which are equipped with β times more energy. N × m × (1- $m_0$) nodes are advanced nodes and equipped with α times more energy as compared to normal nodes [18]. Let initial energy of normal node is $E_0$. Then initial energy each of the advanced and super nodes are $E_0 \times (1+\alpha)$ and $E_0 \times (1+\beta)$ where α, β means that advanced and super nodes have α, β times more energy than the normal nodes. Heterogeneous network total initial energy will be:

$$\begin{aligned} E_{total} &= N.(1-m).E_0 + N.m.(1-m_o).(1+\alpha).E_0 \\ &\quad + N.m.m_0.E_0.(1+\beta) \\ &= N.E_0.(1+m.(\alpha+m_0.(\beta-\alpha))) \end{aligned} \quad (11)$$

Thus due to heterogeneous nodes system has (1 + m. (α + $m_0$. (β – α))) times more energy and epoch of this new system is equal to $\left(\frac{1}{p_{opt}}\right) \times \left(1 + m \times \left(\alpha + m_0(\beta-\alpha)\right)\right)$. The weighed probabilities for various nodes are given by equations (12 – 14) [18].

$$p_i = \begin{cases} \frac{p_{opt}}{1 + m.(\alpha+m_o.(\beta-\alpha))} & (12) \\ \frac{p_{opt}.(1+\alpha)}{1 + m.(\alpha+m_o.(\beta-\alpha))} & (13) \\ \frac{p_{opt}.(1+\beta)}{1 + m.(\alpha+m_o.(\beta-\alpha))} & (14) \end{cases}$$

Threshold for cluster head selection of normal, advanced, super nodes can be calculated by putting above values in Equation 15.

$$T(s_i) = \begin{cases} \dfrac{p_i}{1 - p_i\left(r \bmod \frac{1}{p_i}\right)} \times \dfrac{E_{(r)}}{E_{(i)}} & \text{If } s_i \in G \\ \dfrac{p_i}{1 - p_i\left(r \bmod \frac{1}{p_i}\right)} & \text{If } s_i \in G' \\ \dfrac{p_i}{1 - p_i\left(r \bmod \frac{1}{p_i}\right)} & \text{If } s_i \in G'' \\ 0 & \text{otherwise} \end{cases} \quad (15)$$

Where G, G' and G'' represents set of normal , advanced and super nodes that are not selected as cluster heads within the last $\frac{1}{p_i}$ rounds of the epoch, depending upon whether $s_i$ represents a normal, advanced or super node. $E_{(r)}$, $E_{(i)}$ represents residual and initial energy of a normal node respectively.

Cluster head threshold for normal nodes are multiplied by the ratio of residual and initial energy of the normal node because they have less energy than advanced and super nodes hence they should become cluster head only when they have sufficient remaining energy for performing this duty.

Nearly 70 percent energy of a WSN is consumed in communication and the energy consumed in transmission dominates the total energy consumed for communication. The transmission power grows exponentially with the increase of transmission distance therefore to reduce energy consumption of a WSN multihop short distance communication is desirable M-EECDA introduces a three tier architectures for its normal node to save their energy (Figure 2). In a round if a normal sensor node becomes a cluster head then after collecting the data from its members it aggregates the data and instead of sending the data directly to sink it will try to find an out advanced or super node such that

- Distance between normal cluster head and advanced or super node is less than the distance between normal cluster head and base station.
- Which is not a cluster head in this particular round?

If the normal cluster head is able to find such an advance or super node that is not a cluster head in this round r and also its distance is less than the distance between normal cluster head and base station then normal cluster head instead of sending the data directly to the base station it sends its data to this advance or super node which further send it to the base station. If normal cluster head does not find any such advance or super node that fulfils the above mentioned two conditions then it will send its aggregated data directly to the base station. Thus by introducing multihop architectures or three tier architectures for normal cluster head, M-EECDA has reduced the energy consumption of network.

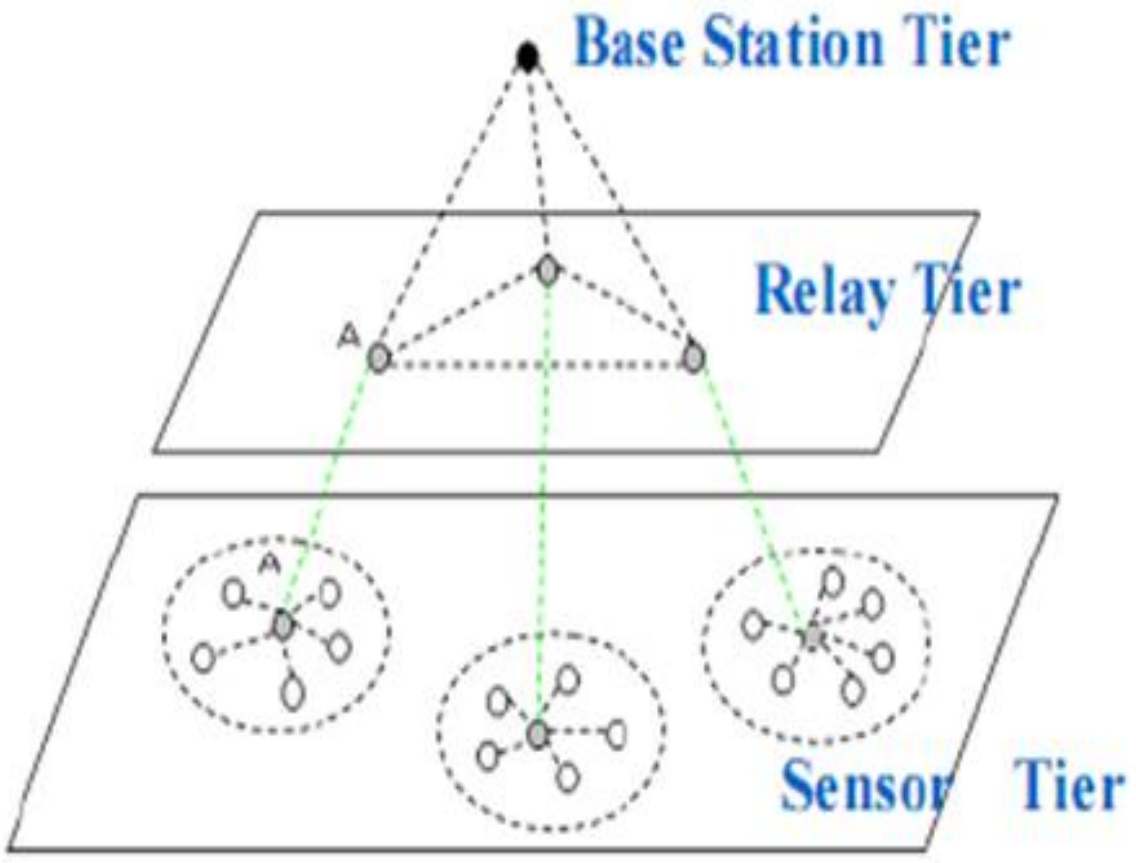

**Figure 2: 3-Tier Architectures for Normal Cluster Heads**

Moreover, when the cluster-heads are selected, each node tries to join the closest (considering the transmission power) cluster-head. However in some cases this is not an optimal choice because, if a sensor node exists in the base station direction whose distance from the base station is less than all the nearby cluster-head distance (see Figure 3).

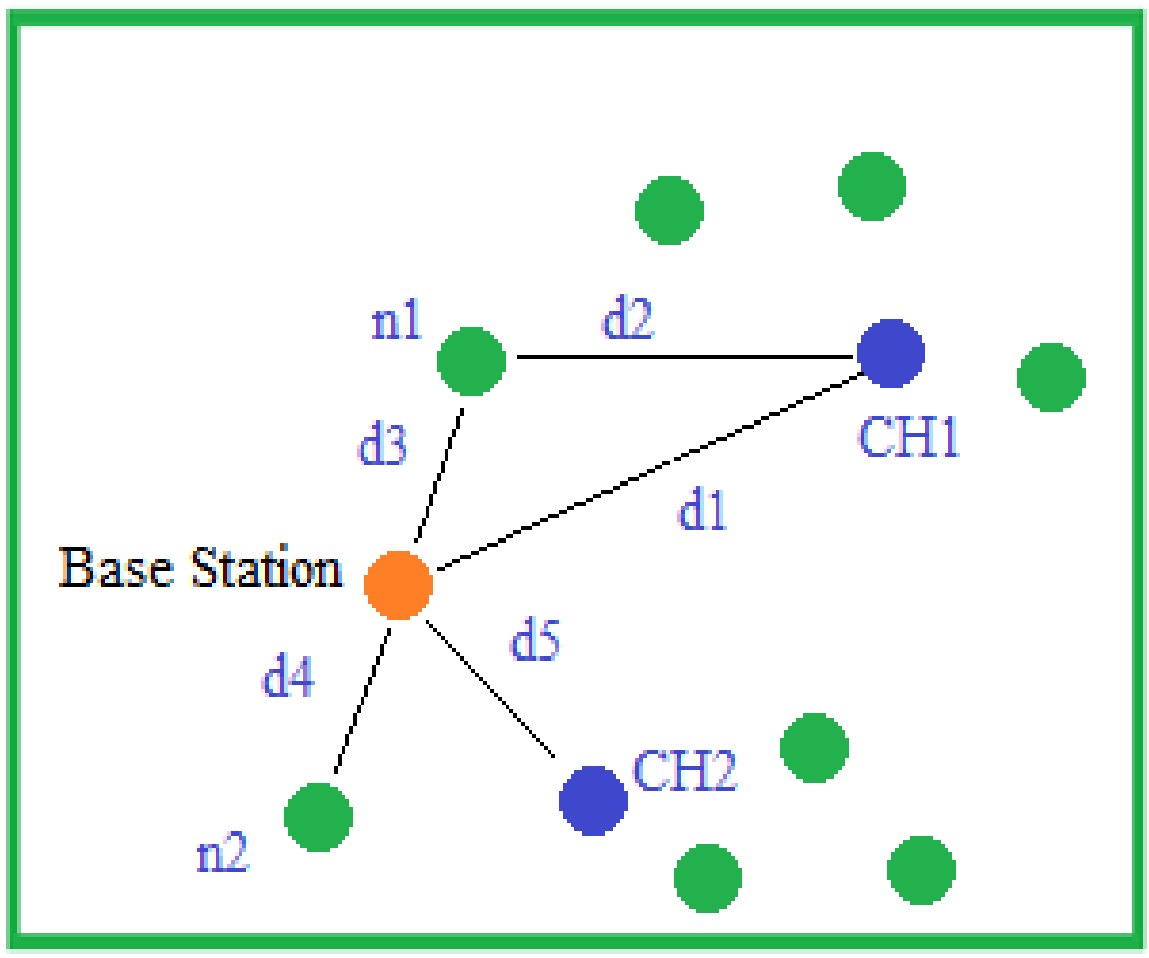

**Figure 3: Cluster Head Selections for Data Transmission**

In Figure 3, node n1 have to transmit an L-bits message to base station. The closest cluster-head to n1 is CH1 and if the node n1 belongs to this cluster it will spend energy (16).

$$E3 = L \cdot E_{elec} + L \cdot \epsilon_{d2} \cdot d_2^u \quad (16)$$

$$\text{Where } \begin{cases} u = 2, \epsilon_{d2} = 10 \text{ pj / bit / m}^2, & if\ d_2 < d_0 \\ u = 4, \epsilon_{d2} = 0.0013 \text{ pj / bit / m}^4, & if\ d_2 \geq d_0 \end{cases} \quad (17)$$

But if the node n1 chooses to transfer data to the base station directly, this energy will be (18):

$$E4 = L \cdot E_{elec} + L \cdot \epsilon_{d3} \cdot d_3^v \quad (18)$$

$$\text{Where } \begin{cases} v = 2, \epsilon_{d3} = 10 \text{ pj / bit / m}^2, & if\ d_3 < d_0 \\ v = 4, \epsilon_{d3} = 0.0013 \text{ pj / bit / m}^4, & if\ d_3 \geq d_0 \end{cases} \quad (19)$$

Here positive coefficients $u\, and\, v$ represent the energy dissipation radio model used. Clearly E4 < E3 but in this case lot of uncompressed data is collected at the base station. To get rid of this problem M-EECDA has introduced a sleep state in the network in the following manner.

When *E3 > E4* it is not an optimal choice for data transmission and energy saving , in this case instead of sending the data to the cluster head *CH1*, sensor node *n1* enters into a sleep state and waits for the next round in which it either itself become a cluster head or finds a nearby cluster head such that *E3 < E4*. Sensor node n1 remains in the sleep state for the maximum 8 rounds, if in these next 8 rounds it either becomes a cluster head or finds a nearby cluster head such that *E3 < E4* it wakes up and performs their respective duty either of a cluster head or the member of a cluster head. If sensor node n1 is neither able to become the member of a cluster head such that *E3 < E4* nor itself become a cluster head in the next 8 rounds , then sensor node wakes up and transmit the data directly to the base station.

**Table 1 Radio Parameters Used in M-EECDA**

| Parameter | Value |
|---|---|
| $E_{elec}$ | 5 nJ/bit |
| $\varepsilon_{fs}$ | 10 pJ/bit/m$^2$ |
| $\varepsilon_{mp}$ | 0.0013 pJ/bit/m$^4$ |
| $E_0$ | 0.5 J |
| $E_{DA}$ | 5 nJ/bit/message |
| Message Size | 4000 bits |
| $p_{opt}$ | 0.1 |
| $d_0$ | 70m |

# 6. SIMULATION AND RESULTS

The performance of M-EECDA is compared with EECDA. For simulation 100 x 100 square meters region with 100 sensor nodes are used as shown in Figure 1. In the Figure normal nodes are denoted by using the symbol (o), advanced nodes with (+), super nodes by (*) and the base station by (x). Radio parameters used for the simulation are given in Table 1. The following performance metrics are used for evaluating the protocol.

(i) **Network Lifetime:** This is the time interval between network start until the death of the last node.

(ii) **Stability Period:** This is the time interval between network start until the death of the first node.

(iii) **Number of Alive Nodes per round:** This will measure the number of live nodes in each round.

(iv) **Number of cluster heads per round:** This will reflect the number of cluster heads formed in each round.

(v) **Numbers of packets send to base station:** This will measure the total number of packets which are sent to base station.

The following cases of heterogeneity are used for evaluating the performance of protocol

Case 1: $m = 0.5$, $m_0 = 0.4$, $a = 1$, $b = 2$

Case 2: $m = 0.5$, $m_0 = 0.4$, $a = 1.5$, $b = 3$

**Case 1: $m = 0.5$, $m_0 = 0.4$, $a = 1$, $b = 2$**

In this case there are 30 advanced nodes having one time and 20 super nodes have two times more energy than normal nodes. Figure 4 shows that network lifetime of M-EECDA is more than EECDA as last node dies in M-EECDA after 11445 rounds and in EECDA it dies after 5369 rounds. Similarly first node dies in M-EECDA after 1625 rounds and in EECDA dies after 1374 rounds. Thus stability period of M-EECDA is more than EECDA. Figure 5 shows that number of alive nodes per round is more in M-EECDA than EECDA. Figure 6 shows that throughput i.e. total number of messages sends to base station is more in M-EECDA. Figure 7 plots the total remaining energy per round and from the figure it is clear that total remaining energy per round is more in M-EECDA than EECDA

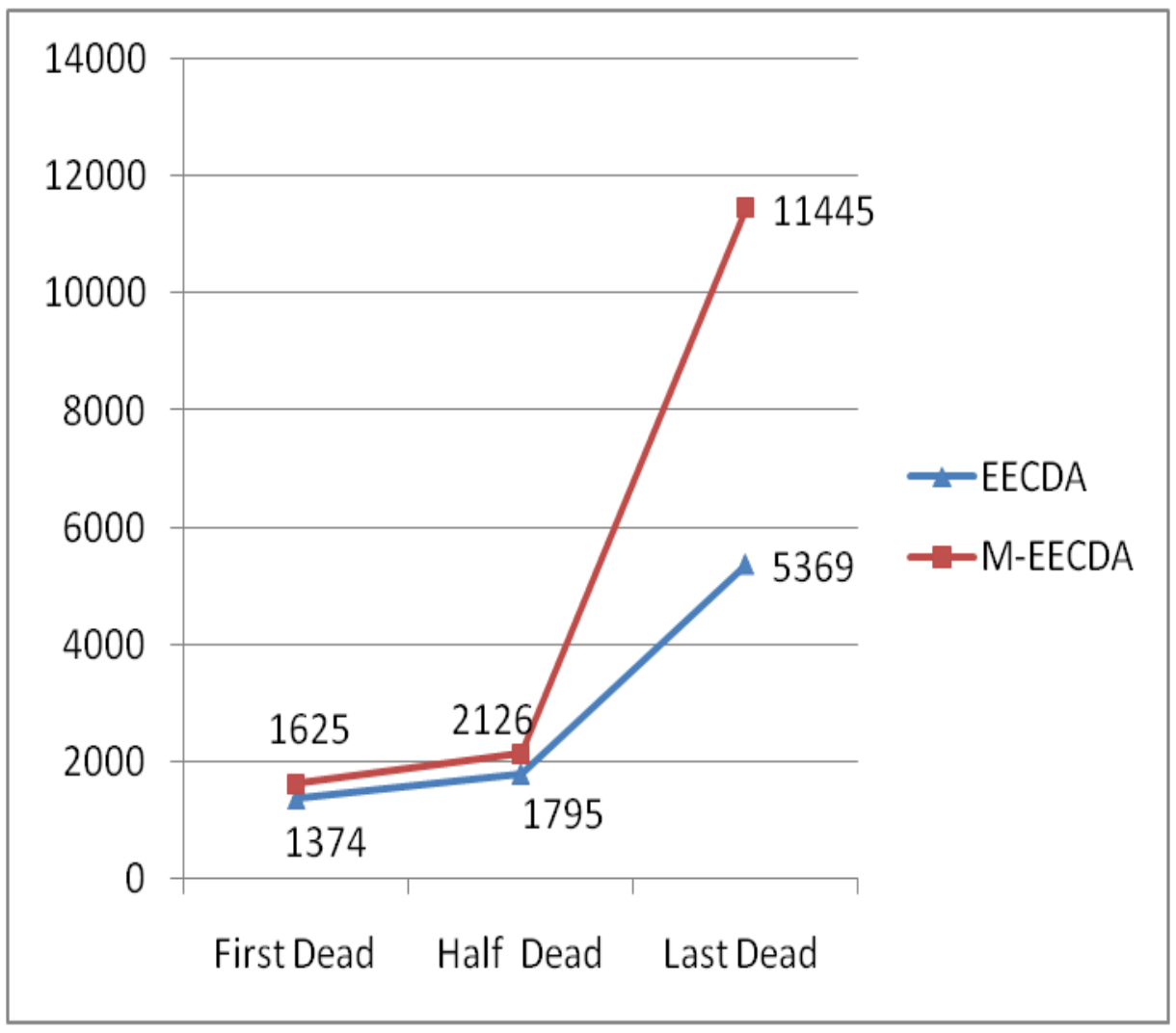

**Figure 4: Rounds for 1st, Half and Last Node Dead in M-EECDA & EECDA**

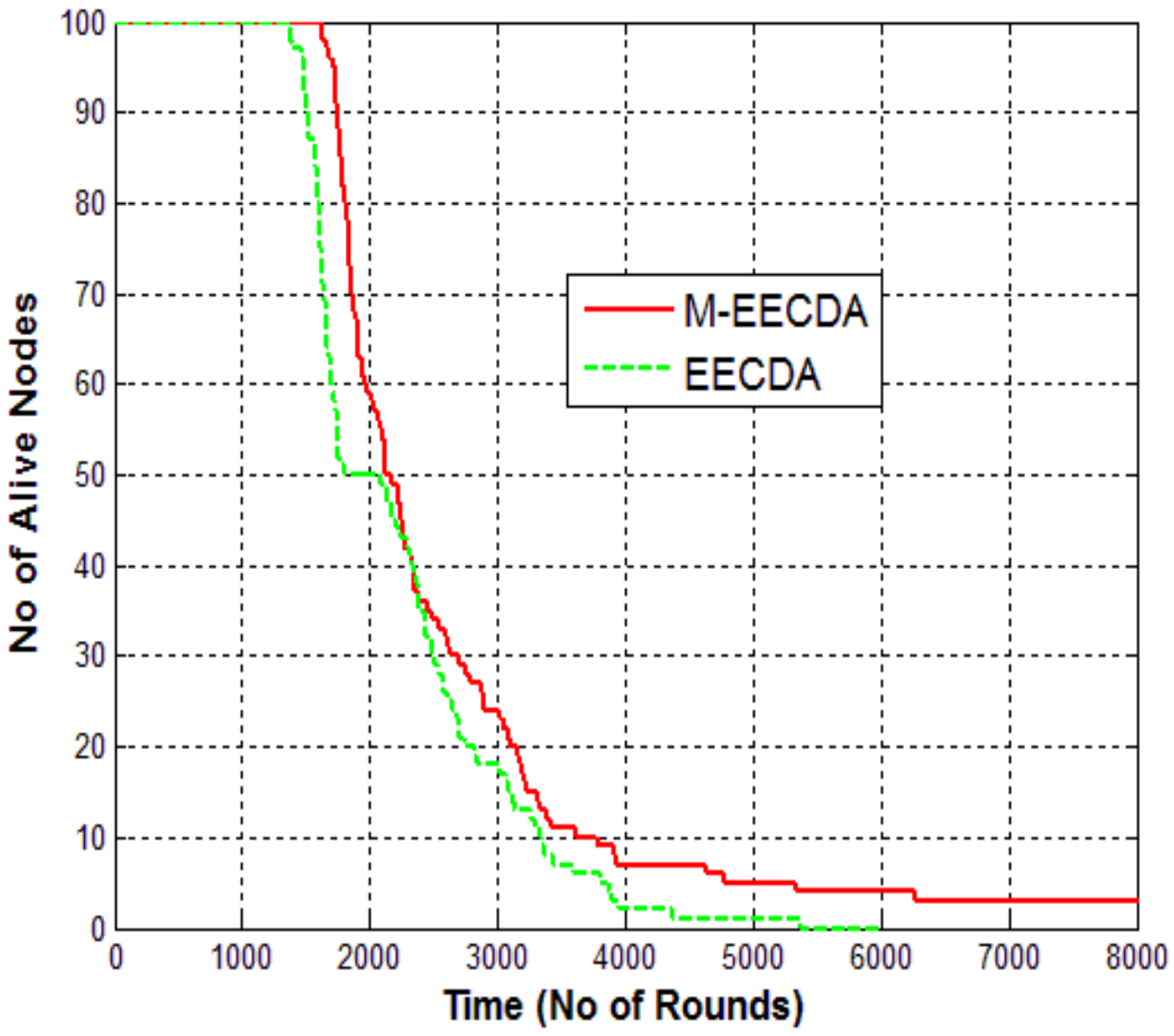

**Figure 5: No. of Alive nodes in M-EECDA & EECDA**

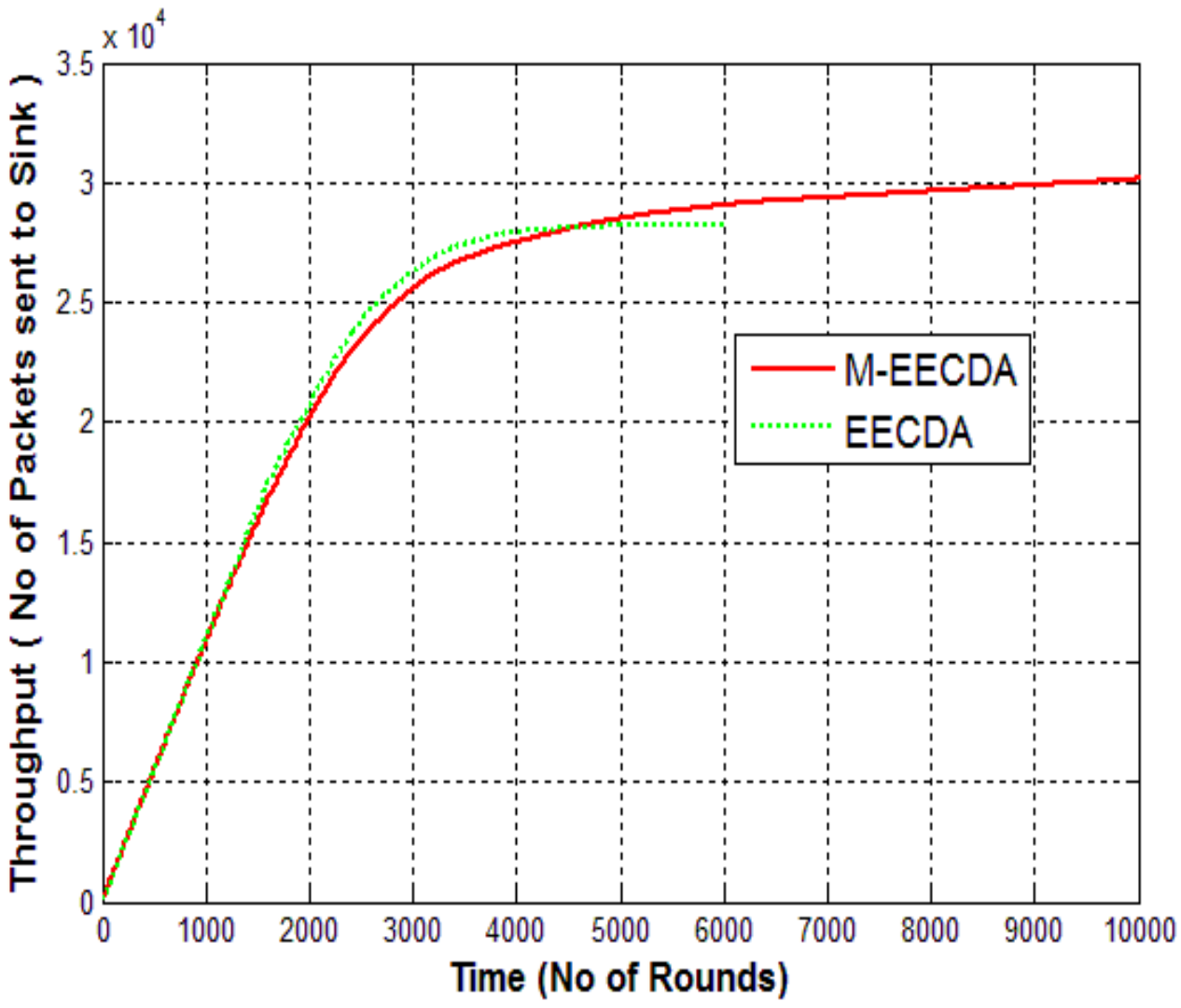

**Figure 6: Throughput in M-EECDA & EECDA**

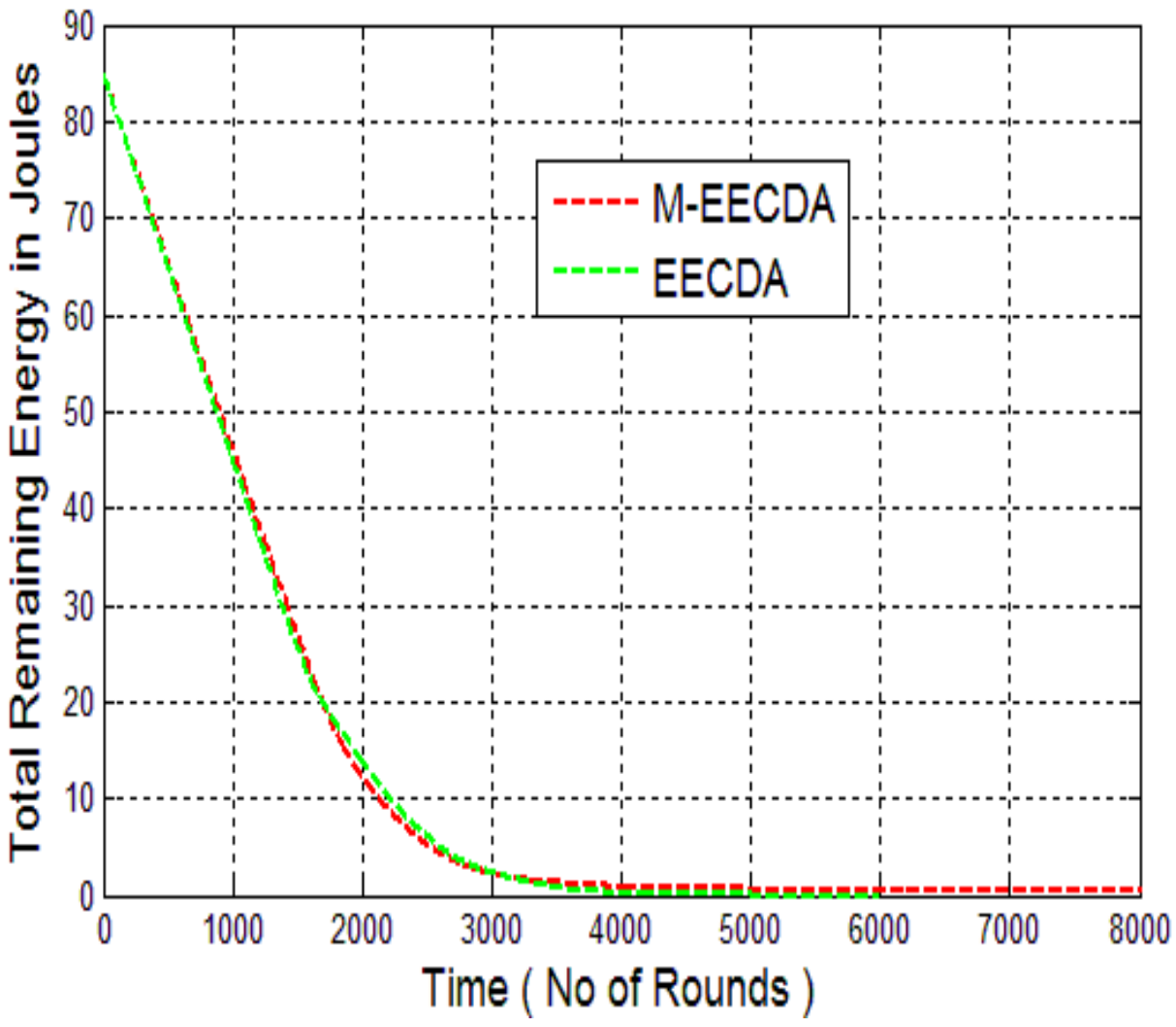

**Figure 7: Total Remaining Energy per round in M-EECDA & EECDA**

**Case 2: m = 0.5, $m_0$ = 0.4, a =1.5, b =3**

In this case there are 30 advanced nodes and 20 super nodes. Advanced nodes have 1.5 times and super nodes have three times more energy as compared to normal nodes. Figure 8 shows that network lifetime of M-EECDA is more than EECDA as last node dies earlier in EECDA. In M-EECDA last node dies after 11995 rounds and in EECDA dies after 7125 rounds. In M-EECDA and EECDA first node dies after 1705 and 1480 rounds respectively. Thus stability period of M-EECDA is more than EECDA. Figure 9 shows that no of alive nodes per round are more in M-EECDA than EECDA. Figure 10 shows that throughput i.e. total number of packets sends to base station is more in M-EECDA. Figure 11 plots the total remaining energy per round and total remaining energy per round is more in M-EECDA as compared to EECDA.

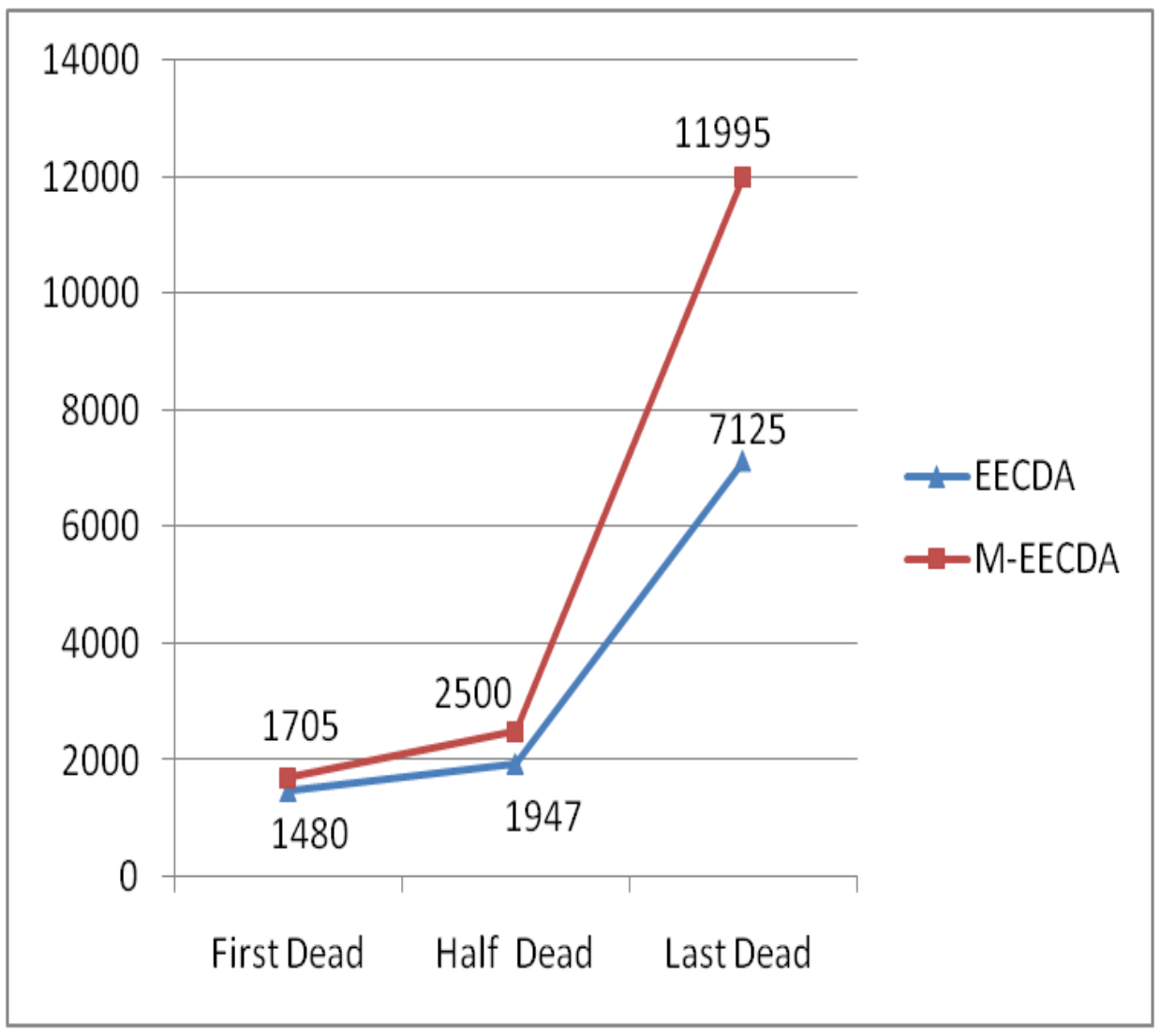

**Figure 8: Rounds for 1st, Half and Last Node Dead in M-EECDA & EECDA**

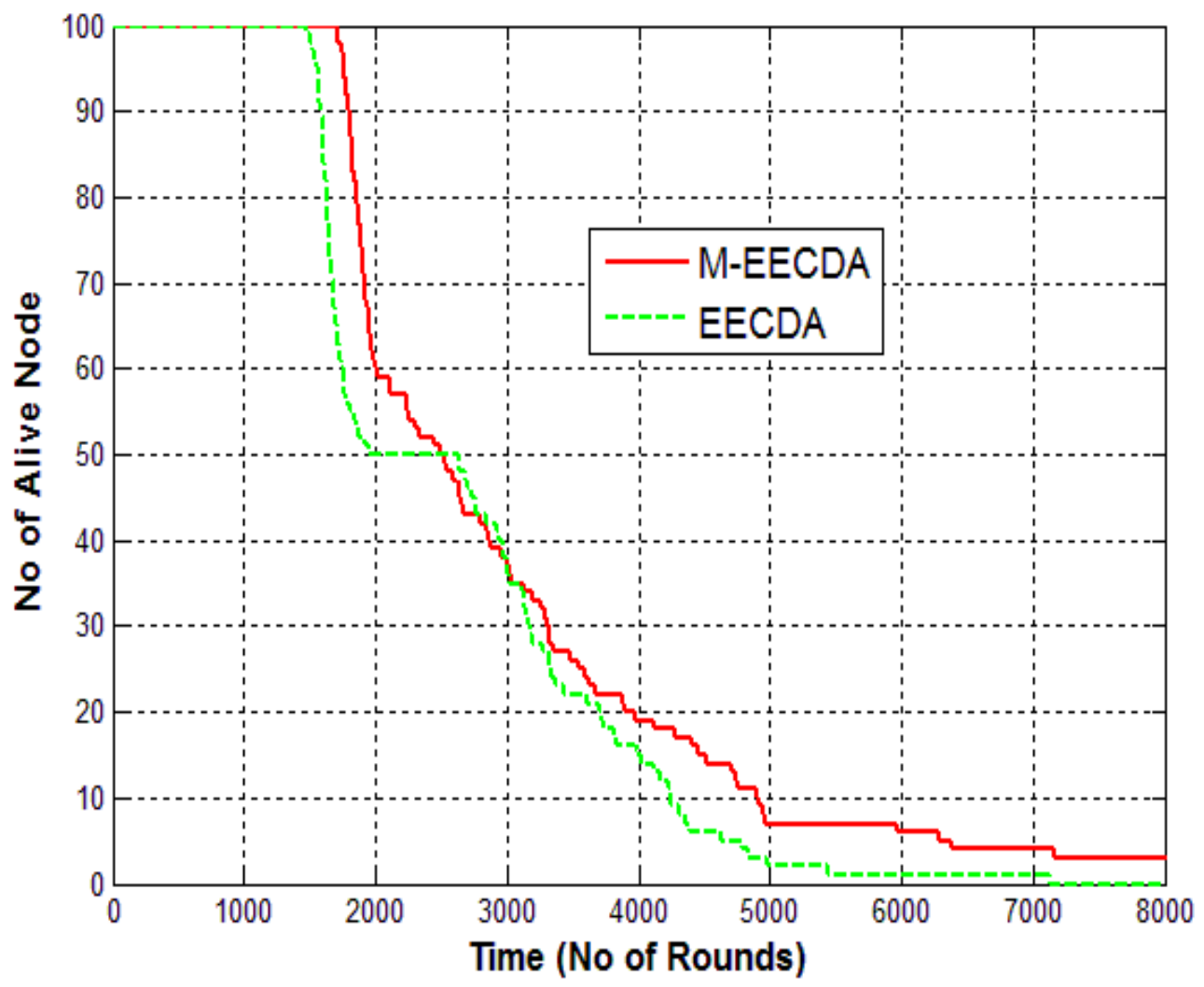

**Figure 9: No. of Alive Nodes in M-EECDA & EECDA**

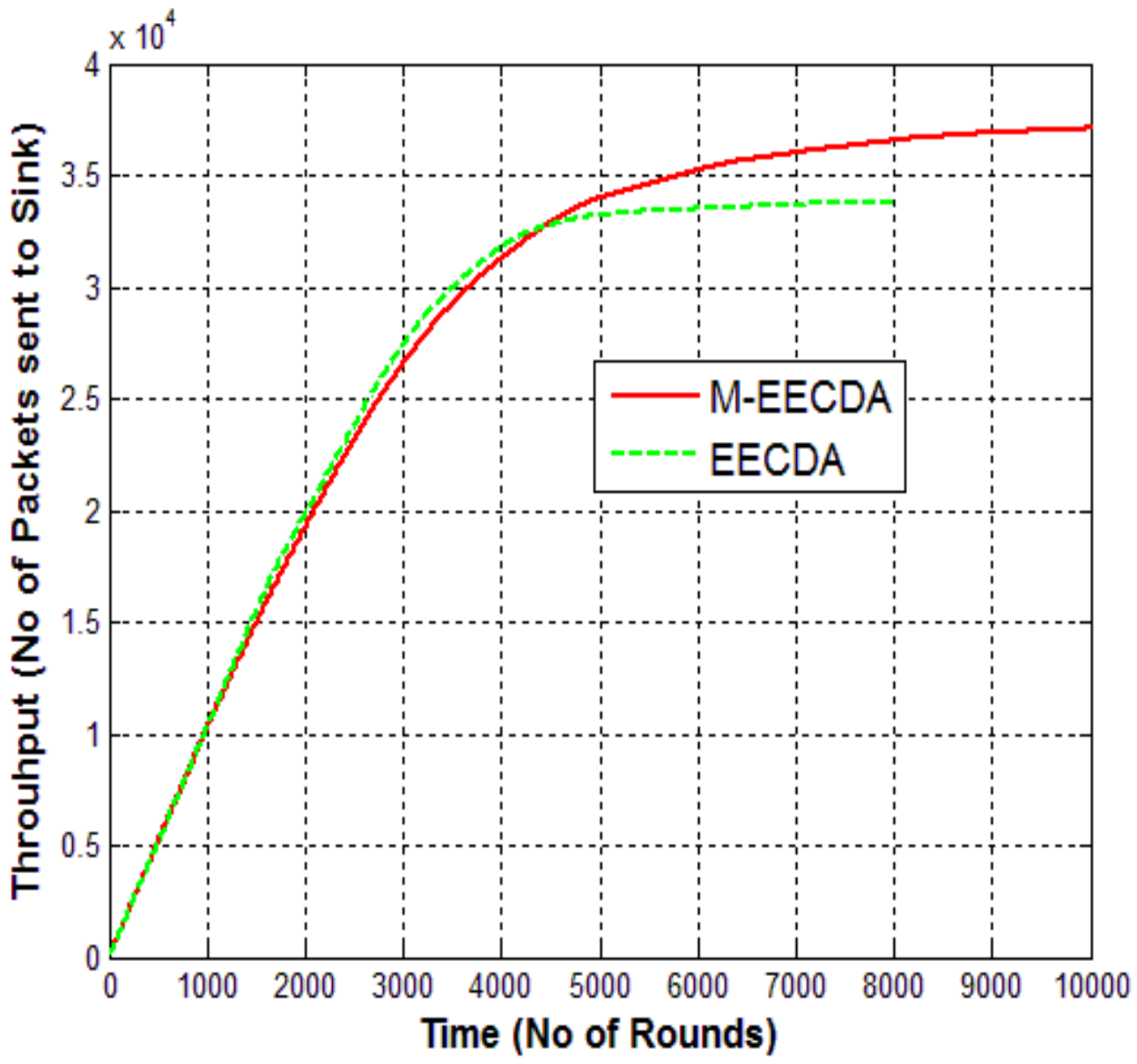

**Figure 10: Throughput in M-EECDA & EECDA**

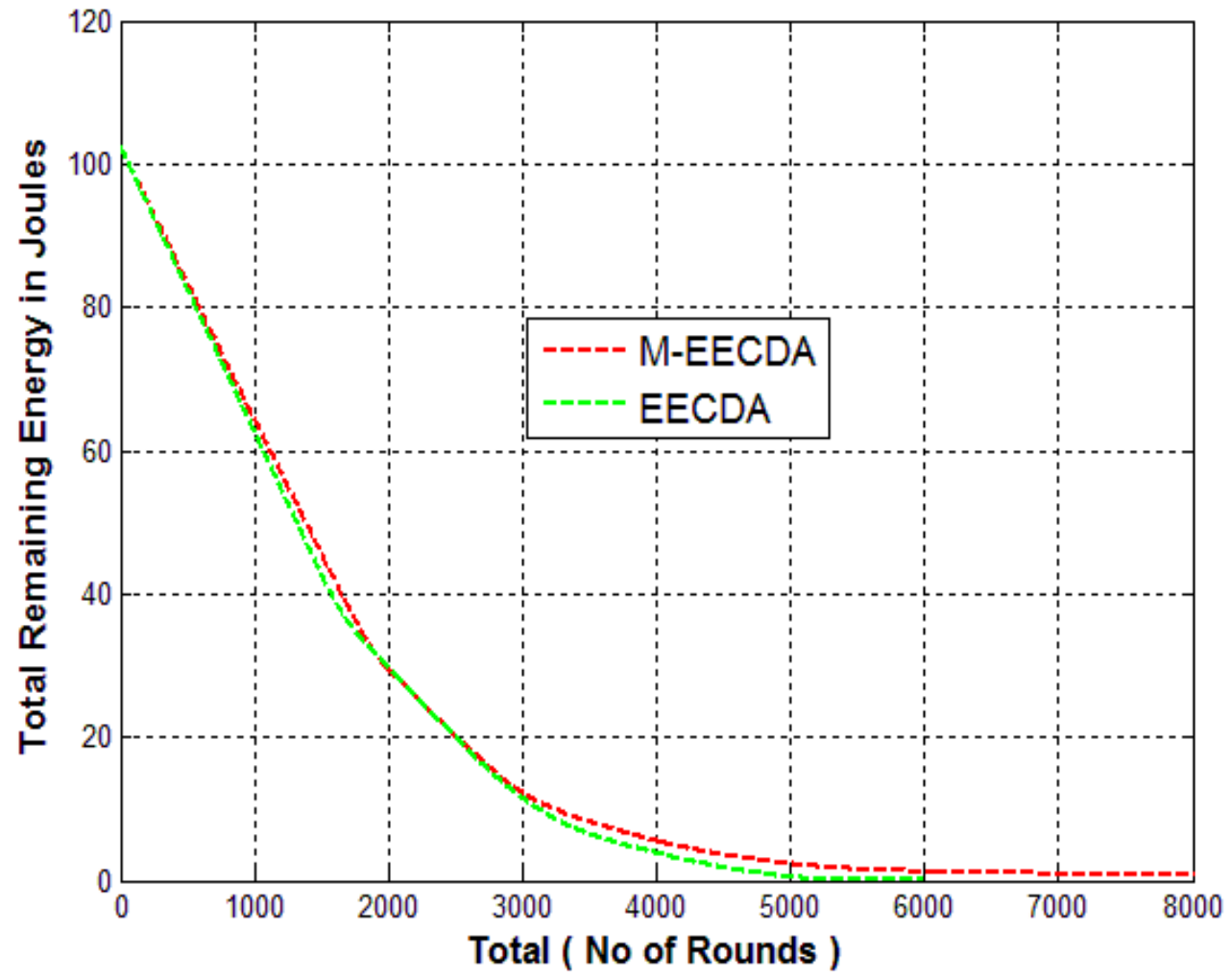

**Figure 11: Total Remaining Energy per Round in M-EECDA & EECDA**

## 7. CONCLUSION

This paper has described M-EECDA a multihop energy efficient communication protocol for three level heterogeneous networks which takes the full advantage of heterogeneity. It improves the network lifetime, stable region and throughput of the network. For increasing energy efficiency of network it uses a residual energy based cluster head election scheme for normal nodes. Further advance and super nodes act as the relay node for transmitting the data load of normal cluster head when they are not performing the duty of a cluster head. Simulation result shows that it is better than the existing EECDA protocol for three level heterogeneous networks.